\documentclass[superscriptaddress,preprintnumbers,nofootinbib,prd]{revtex4}
\usepackage{longtable}
\usepackage{amsmath,amssymb}
\usepackage{graphicx}
\usepackage{dcolumn}
\usepackage{epsfig,color}
\usepackage{bm}
\usepackage{verbatim}
\usepackage{bbold}
\usepackage{rotating}

\newcommand{\nn}{\nonumber}
\def\ee{\end{equation}}
\def\be{\begin{equation}}

\newcommand{\bnabla} {{\mbox{\boldmath $\nabla$}}}
\newcommand{\bdel} {{\mbox{\boldmath $\nabla$}}}
\newcommand{\bPi}{{\mbox{\boldmath $\Pi$}}} 
\def\A{{\bf A}}  
\def\B{{\bf B}}
\def\x{{\bf x}}
\def\r{{\bf r}}
\def\y{{\bf y}} 
\def\k{{\bf k}}

\def\q{{\bf q}}

\def\D{{\bf D}}

\begin{document}

\title{Heavy Hybrid Decays in a Constituent Gluon Model}
\author{Christian Farina}
\affiliation{Department of Physics and Astronomy, University of Pittsburgh, Pittsburgh PA 15260}
\author{Hugo Garcia Tecocoatzi}
\affiliation{Department of Physics, University of La Plata (UNLP), 49 y 115 cc. 67,
1900 La Plata, Argentina}
\author{Alessandro Giachino}
\affiliation{INFN, Sezione di Genova, via Dodecaneso 33, 16146 Genova, Italy}
\author{Elena Santopinto}
\affiliation{INFN, Sezione di Genova, via Dodecaneso 33, 16146 Genova, Italy}
\author{Eric S. Swanson}\email[]{swansone@pitt.edu}
\affiliation{Department of Physics and Astronomy, University of Pittsburgh, Pittsburgh PA 15260}

\begin{abstract}
A constituent gluon model that is informed by recent lattice field theory is developed. The model is then used to compute hybrid strong decay widths, that can be useful for the GlueX collaboration at Jefferson Lab and the PANDA collaboration at FAIR. Commensurately, forthcoming data from GlueX and PANDA will test the model. Widths tend to be typical of charmonia, except for those of the lightest hybrid $S$-wave multiplet. Selection rules, extensions, limitations, and applications are discussed.
\end{abstract}

\maketitle

\section{Introduction}

Longstanding interest in hybrid mesons continues because they offer a window into the unknown dynamics of nonperturbative gluonic degrees of freedom in Quantum Chromodynamics (QCD). In particular, robust experimental programs are underway by the GlueX collaboration at Jefferson Lab\cite{Dobbs} and by the PANDA collaboration at FAIR\cite{Lutz:2009ff}. The spectrum, radiative transitions, and strong decay widths are of interest to the experimental community as these are crucial to informing experimental design and the interpretation of experimental results. A simple model of hybrid structure is developed in the following and is applied to the strong decays of charmonium hybrids, which is a focal point for the PANDA collaboration.

A variety of hybrid models have been developed since the 1970s; these chiefly fall into two camps: constituent gluon models and flux tube models. The former include bag models\cite{bags} or simple constituent glue models\cite{HM}, while the latter include string models\cite{string} or more elaborate flux tube models\cite{flux}. Most of this work predates the substantial progress that has been made in understanding hybrid properties by the lattice gauge community.
This effort had revealed flaws in the early models, most of which are severe enough to invalidate the models
(see Ref. \cite{Meyer:2015eta} for a recent review of the experimental and theoretical status of hybrids). Given this situation, and the impending data from GlueX and PANDA, it is incumbent to revisit hybrid structure modelling. Of course, it is preferable that new models incorporate the features of gluodynamics that the past 20 years of lattice gauge computations have revealed. Among these are the spectrum of adiabatic gluonic excitations\cite{Juge:1997nc,Juge:1999ie}, the gluelump (bound states of gluons and a static adjoint colour source) spectrum\cite{gluelump}, and properties of charmonium hybrids\cite{Liu:2012ze,Cheung:2016bym,Knechtli:2019bqx}. Of particular interest is the confirmation that the heavy quark multiplet structure anticipated in Ref. \cite{Juge:1999ie} is reflected in the charmonium spectrum\cite{Liu:2012ze}. It is interesting, and very suggestive, that this multiplet structure can be reproduced by degrees of freedom consisting of a quark, an antiquark, and an axial gluon with quantum numbers $J^{PC} = 1^{+-}$\cite{Guo:2008yz,Dudek:2011bn}.  

Indeed, work has started along these lines, starting with an examination of the adiabatic gluon potentials in a constituent gluon model based on the Hamiltonian of QCD\cite{Swanson:1998kx}. This work obtained decent agreement with lattice but noted that the level ordering was incorrect. Subsequently it was realized that gluonic three-body interactions can correct the level ordering problem\cite{Szczepaniak:2006nx}, and this was used to construct models for gluelumps\cite{Guo:2007sm} and heavy quark hybrids\cite{Guo:2008yz}. Here we shall follow a similar approach with an eye to developing the simplest model possible that captures the necessary features revealed by the lattice. This will be described in Section \ref{sec:model}. The resulting hybrid wavefunctions will then form the starting point for the computation of strong decays of charmonium hybrids.

It is useful to note that advances in the application of effective field theory to hybrids have also been occurring. This approach has led to a more thorough understanding of systematics in heavy quark hybrids, including subtleties concerning angular momentum and the structure of subleading spin-dependent operators\cite{Berwein:2015vca, Brambilla:2018pyn}. Although these results are interesting, and inform our model building, they are not of direct relevance to the present work because gluonic degrees of freedom are integrated out, precluding the application of the formalism to hadronic transitions.

Hybrid meson decay models also have a long history. The earliest work we are aware of is due to Tanimoto\cite{Tanimoto} (see also the first of Ref.\cite{HM}), who assumed a simple $P$-wave symmetric Gaussian wavefunction for all hybrids and decay via gluon dissociation to a quark-antiquark pair. In contrast, the model of Isgur and Paton assumed that gluon dynamics can be described by a nonrelativistic string\cite{Isgur:1984bm} and that strong decays occur via a string breaking mechanism with vacuum quantum numbers\cite{Isgur:1985vy}. A model for hybrid decay that employed the same philosophy but assumed vector quark pair creation was developed by Swanson and Szczepaniak\cite{Swanson:1997wy} and applied in Ref. \cite{Page:1998gz}. 

The approach adopted here is predicated on the simple constituent gluon model we adopt and is guided by the QCD Hamiltonian. Namely, we shall assume that the gluon simply dissociates into a light quark pair via the perturbative quark-gluon coupling. Thus we resurrect the Tanimoto decay model, but with the addition of a reasonable model of hybrid structure and with the benefit of lattice predictions of the charmonium hybrid spectrum. The decay model will be described in Section \ref{sec:decay} followed by conclusions in Section \ref{sec:conc}.

\section{Constituent Gluon Model of Hybrid Mesons}
\label{sec:model}

Our model for hybrid structure will assume constituent quarks and gluons are the dominant degrees of freedom and that these interact according to QCD. Interactions are made especially transparent by adopting the Hamiltonian of QCD in Coulomb gauge\cite{Schwinger, Christ:1980ku}. Perhaps the strongest reason for adopting this approach is that all degrees of freedom are manifest (\textit{i.e.}, there are no ghosts) and that an explicit instantaneous ``Coulomb" interaction exists between valence quarks and gluons. The philosophy advocated here was developed extensively in Ref. \cite{Szczepaniak:2001rg}, where a mean field model of the gluonic vacuum was used to derive an effective gluon dispersion and a nonperturbative expression for the Coulomb operator (to be explained further below) that recapitulates linear confinement as observed in lattice gauge computations.

The construction of a hybrid meson proceeds by assuming a dynamical constituent gluon (thus, the gluon remains transverse) coupled to quarks. This involves a Clebsch-Gordan coefficient that couples the canonical gluon spin projection $s$ to the gluon angular momentum $(\ell_g,m_g)$ to total gluon spin, $j_g$. Converting to the gluon helicity basis and assuming  that  $\ell_g = j_g$ reduces the product of two Wigner matrices to one and produces a factor of

\be
\chi^{(-)}_{\lambda,\mu} \equiv \langle 1 \lambda \ell_g 0| \ell_g \mu\rangle = 
\begin{cases} 0, \ell_g = 0 \\ \frac{\lambda}{\sqrt{2}} \delta_{\lambda,\mu}, \ell_g \geq 1 \end{cases}.
\ee
This represents a transverse electric (TE) gluon in our model and forms the explicit realisation of the axial constituent gluon. Alternatively, one may set $\ell_g = j_g\pm 1$ and obtain a transverse magnetic (TM) gluon with a Clebsch factor given by

\be
\chi^{(+)}_{\lambda,\mu} = \frac{1}{\sqrt{2}} \delta_{\lambda,\mu}.
\ee

The end result for a generic hybrid meson creation operator is then

\begin{align}
&|JM [LS \ell j_g \xi]\rangle = \frac{1}{2} T_{ij}^A\int \frac{d^3q}{(2\pi)^3}\, \frac{d^3k}{(2\pi)^3} \,
\Psi_{j_g,\ell}({\bf k}, {\bf q})\, \sqrt{\frac{2 j_g +1}{4\pi}} \, D_{m_g\mu}^{j_g*}(\hat k) \, \chi^{(\xi)}_{\mu,\lambda} \nn \\
& \times 
\langle \frac{1}{2} m \frac{1}{2} \bar{m} | S M_{S} \rangle \,
\langle \ell m_\ell, j_g m_g| L M_L\rangle \, 
\langle S M_S, L M_{L} | J M \rangle \,
b_{{\bf q}-\frac{{\bf k}}{2},i,m}^\dagger \,
  d_{-{\bf q}-\frac{{\bf k}}{2},j, \bar{m}}^\dagger \,
a^\dagger_{{\bf k}, A, \lambda} |0\rangle,
\label{eq:PSI}
\end{align} 
The momenta of the constituents are chosen as a convenient rescaling of Jacobi coordinates, 
$\vec k = -2/\sqrt{6} \vec p_\lambda$ and $ \vec q = \vec p_\rho/\sqrt{2}$.
Finally, the hybrid state is an eigenstate of parity and charge conjugation with eigenvalues given by

\be
P=\xi (-1)^{\ell + j_g+1}
\ee
and
\be
  C= (-1)^{\ell+S+1}.
\ee

As mentioned above, we choose to model hybrid structure with a simplified version of the QCD Hamiltonian. The Hamiltonian in Coulomb gauge is written as\cite{Schwinger, Christ:1980ku}

\begin{equation}
H_{QCD} = \int d^3x\, \left[ \psi^\dagger\left( -i \alpha\cdot\bdel
 + \beta m\right) \psi + 
\frac{1}{2}\left( {\cal J}^{-1/2}\bPi {\cal J} 
\cdot \bPi {\cal J}^{-1/2} + \B\cdot\B\right) 
 -g \psi^\dagger \bm{\alpha}\cdot \A \psi \right] + H_C
\label{eq:h}
\end{equation}
with

\begin{equation}
H_C = \frac{1}{2}\int d^3x\, d^3y\, {\cal J}^{-1/2} \rho^A(\x) 
 {\cal J}^{1/2} \hat K_{AB}(\x,\y;\A) {\cal J}^{1/2} \rho^B(\y) {\cal J}^{-1/2}.
\label{eq:hc}
\end{equation}
The latter expression is the instantaneous interaction that emerges from the imposition of Gauss's law in Coulomb gauge. This involves the Faddeev-Popov determinant

\be
{\cal J} \equiv {\rm det}(\bnabla\cdot \D),
\ee
which is the remnant of the ghost in this gauge. Here $D$ represents the adjoint covariant derivative

\be
 \D^{AB} \equiv \delta^{AB} \bnabla  - g f^{ABC}\A^C.
\ee

The colour charge density that appears in Eq. \ref{eq:hc} is given by

\begin{equation}
\rho^A({\bf x}) = 
 f^{ABC} {\bf A}^B({\bf x}) \cdot {\bf \Pi}^C({\bf x}) + \psi^{\dag}(\x)T^A\psi(\x).
\label{eq:rho}
\end{equation}

The kernel of the Coulomb interaction can be formally written as\cite{Christ:1980ku}

\begin{equation}
\hat K^{AB}({\bf x},{\bf y};\A) \equiv \langle{\bf x},A|
 \frac{ g }{ \bnabla\cdot {\bf D}}(-\bnabla^2)
 \frac{ g }{ \bnabla\cdot {\bf D}}|{\bf y},B\rangle.
\label{eq:K}
\end{equation}
Finally, $\A$ is the vector potential and ${\bm \Pi}$ is the conjugate momentum given by the negative of the transverse chromoelectric field.

Equations \ref{eq:h} to \ref{eq:K} are a full field-theoretic representation of QCD and are therefore difficult to solve. In particular, the Coulomb interaction involves infinitely many gluons that build $n$-body operators in Fock space. The two-body operator can be related to the Wilson loop linear confinement potential in a rather direct way\cite{Szczepaniak:2001rg, Zwanziger:2002sh}. Here we accept the mapping to linear confinement as a phenomenologically useful device and simply set 

\be
\hat K^{AB}(\r,0) \to
K^{AB}({\bm r},0) = \delta^{AB}\left( \frac{a_S}{r} - \frac{3}{4} b r\right).
\label{eq:V}
\ee
Of course, this reproduces the successes of the Cornell potential in nonrelativistic quark models. Higher terms in the $n$-body expansion of $\hat K$ can be incorporated in the formalism as required. The effects of the Faddeev-Popov determinant are assumed to be largely confined to restricting the gauge field to the fundamental modular region\cite{Greensite:2011zz}, and are therefore ignored in the following. The vector potential is approximated by its Abelian analogue when making field expansions:

\be
\A^B(\x) = \int \frac{d^3k}{(2\pi)^3}\, \frac{1}{\sqrt{2\omega(k)}}\left( {\bm a}^B(\k) + {\bm a}^{B\dagger}(-\k)\right) {\rm e}^{i \k \cdot \x}.
\label{eq:A}
\ee
The gluon dispersion is denoted as $\omega$ in this expression and need not take on the perturbative form $\omega = k$ because the gluon is a quasiparticle in this approach. In fact, explicit computation in a mean field vacuum model gives a dispersion that is well approximated by\cite{Szczepaniak:2001rg}

\be
\omega^2 = k^2 + m_g^2{\rm e}^{-k/b_g}
\label{eq:omega}
\ee
where the dynamical gluon mass is $m_g \approx 600$ MeV and the parameter $b_g \approx 6000$ MeV.
Finally, the nonrelativistic limit is taken at all opportunities.

The resulting model can be thought of as a minimal extension of the nonrelativistic potential quark model with the addition of constituent gluon degrees of freedom and possible many-body couplings (for example, the tri-linear gluon coupling). It can also be extended to include relativistic effects in the inverse mass expansion that may be important for spin-dependent effects.

Our immediate goal is to obtain a rough description of the lowest hybrid charmonium mesons. As mentioned above, these fall into multiplets that are associated with interpolating fields as shown in Table \ref{tab:JPCMulti}. A model description of these multiplets is sought by employing a TE constituent gluon with $j_g=1$. This leaves the quark angular momentum ($\ell$) and the total angular momentum ($L$) to be specified. Doing so gives columns three to six in the table. Finally combining these degrees of freedom with the total quark spin gives the quantum numbers shown in the last column (the entries in brackets correspond to $S=1$).

\begin{table}[h]
\caption{$J^{PC}$ Hybrid Multiplets. With generating operators and corresponding constituent gluon model quantum numbers.}
\begin{tabular}{c|c|cccc|l}
\hline\hline
multiplet & operator & $\xi$ & $j_g$ & $\ell$ & $L$ & $J^{PC}$ \\
\hline
$H_1$ & $\psi^\dagger \vec B \chi$ & -1 & 1 & 0 & 1 & $1^{--}$, $(0,1,2)^{-+}$ \\
$H_2$ & $\psi^\dagger \nabla \times \vec B \chi$  & -1 & 1 & 1 & 1 & $1^{++}$, $(0,1,2)^{+-}$ \\
$H_3$ &$\psi^\dagger \nabla \cdot \vec B \chi$ & -1 & 1 & 1 & 0 & $0^{++}$, $(1^{+-})$ \\
$H_4$ & $\psi^\dagger [\nabla \vec B]_2 \chi$ & -1 & 1 & 1 & 2 & $2^{++}$, $(1,2,3)^{+-}$ \\
\hline\hline
\end{tabular}
\label{tab:JPCMulti}
\end{table}

In this preliminary calculation we choose a simple product Ansatz for the hybrid wavefunction, namely

\be
\Psi_{j_g,\ell}({\bf k}, {\bf q}) = \varphi_{j_g}(k;\beta)\, \varphi_{\ell}(q; \alpha) Y_{\ell,m_\ell}(\hat q)
\label{eq:ansatz}
\ee
This makes explicit the angular momentum dependence in the $q$ coordinate. The gluon angular momentum dependence is contained  in the Wigner rotation matrix in Eq. \ref{eq:PSI}. In the following we shall employ simple harmonic oscillator (SHO) wavefunctions for the $\varphi$ functions and determine the width parameters  $\alpha$ and $\beta$ variationally:

\be
\frac{\partial}{\partial \alpha(\beta)} \langle J'M'[L' S' \ell' j_g' \xi']|H_{QCD}|JM[L S \ell j_g \xi]\rangle =0.
\label{eq:var}
\ee

\begin{figure}[ht]
\includegraphics[width=0.3\linewidth]{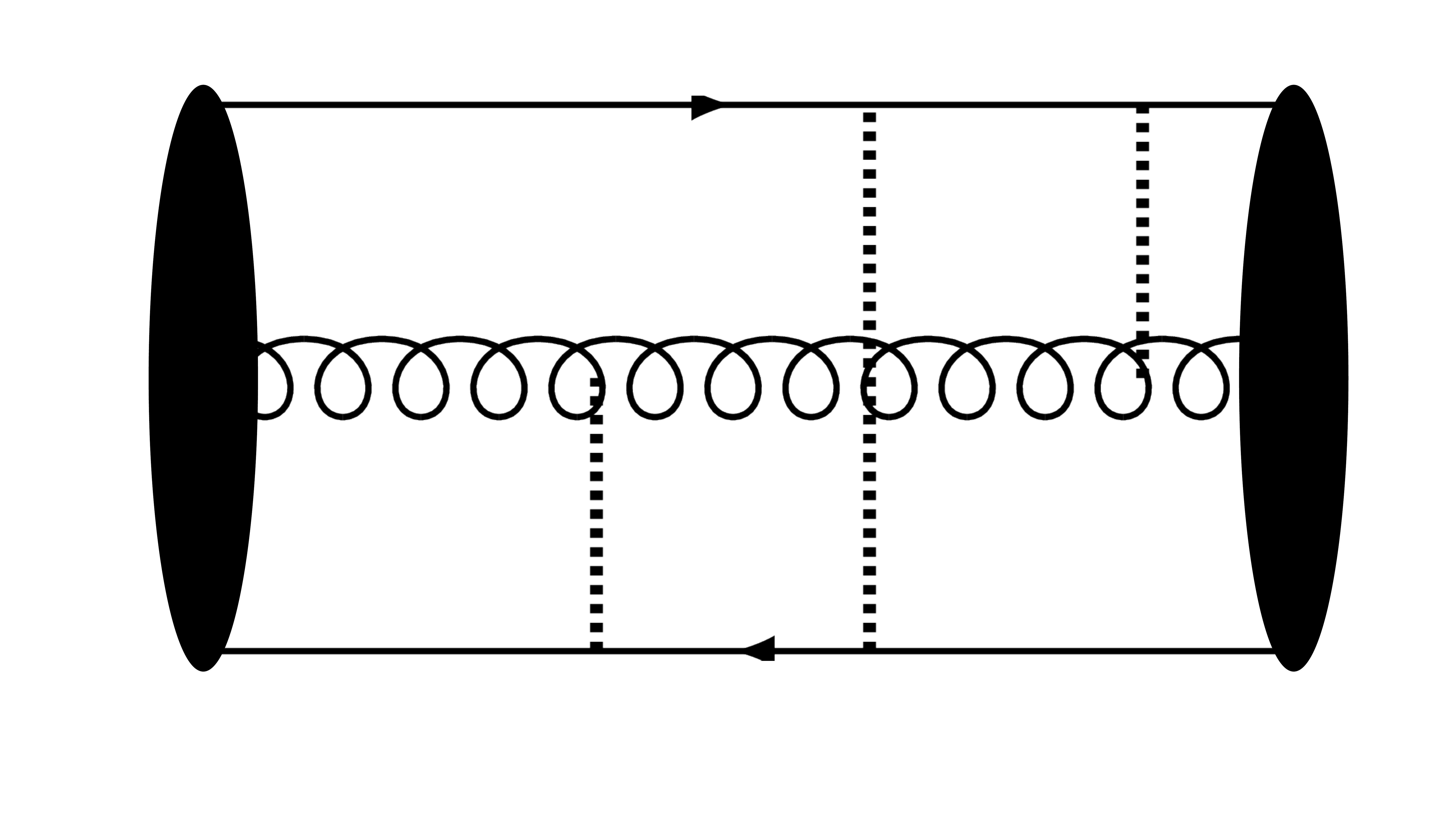}
\caption{Interactions Contributing the Hybrid Bound State.}
\label{fig:Hint}
\end{figure}

The computation amounts to evaluating the diagram illustrated in Fig. \ref{fig:Hint}, where three possible two-body interactions are shown.   As mentioned in the Introduction, it has been found that three-body interactions are important for describing higher gluonic excitation surfaces. However, these are zero for the lowest lying surface\cite{Szczepaniak:2006nx}, with which we deal exclusively.

\begin{figure}[ht]
\includegraphics[width=0.75\linewidth]{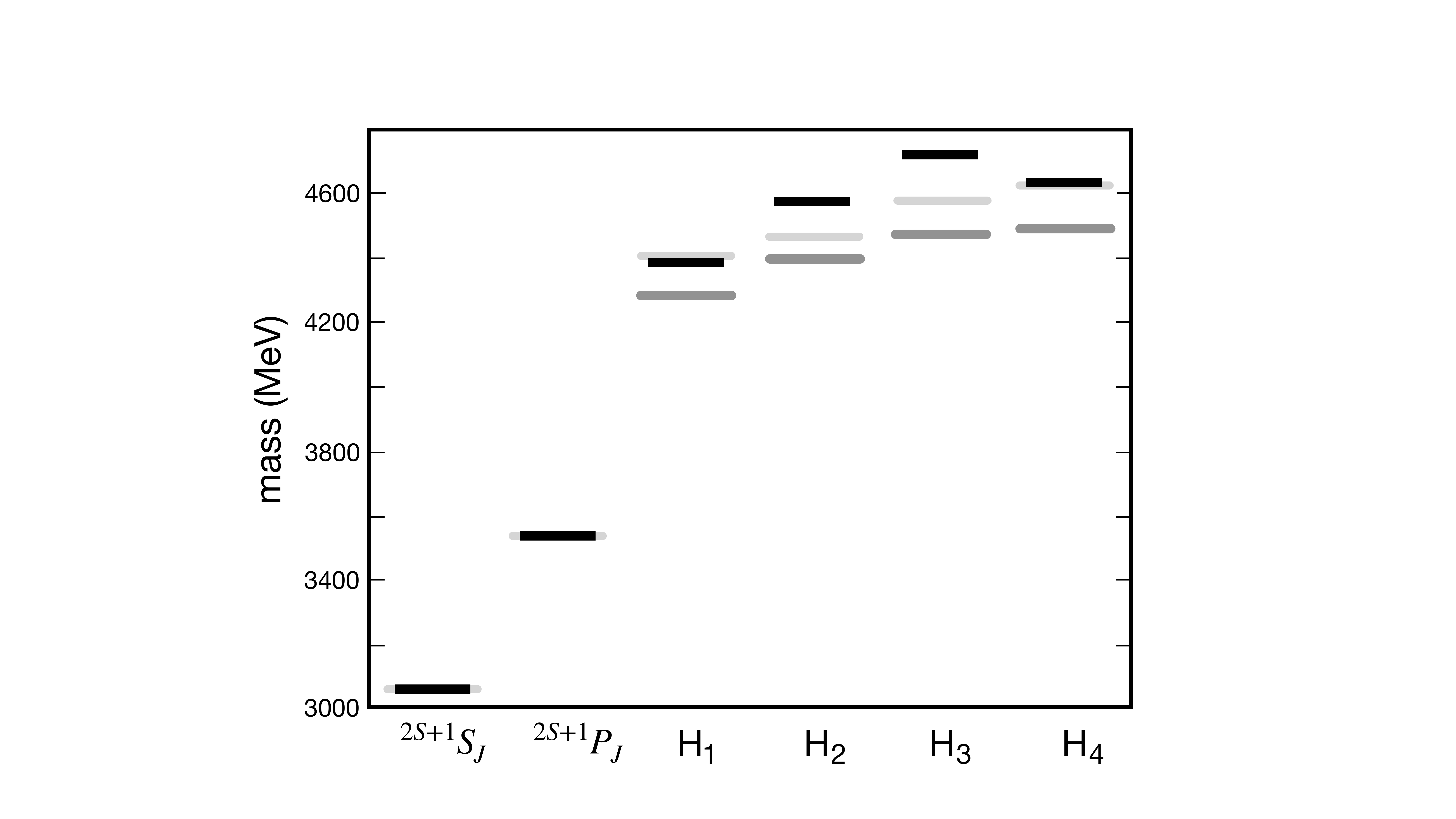}
\caption{Spin Averaged Meson and Hybrid Masses. Dark grey boxes: lattice results with $m_\pi$ = 400 MeV\cite{Liu:2012ze}. 
Light grey boxes: lattice results with $m_\pi$ = 236 MeV\cite{Cheung:2016bym}.  Solid box: model results.}
\label{fig:Hmasses}
\end{figure}

Model parameters were fixed to those used previously in a potential quark model\cite{Swanson:1992ec} along with the dispersion of Eq. \ref{eq:omega}. Thus all parameters have been fixed by previous work. Since this preliminary work is not concerned with fine structure in the spectrum we restrict attention to the nonrelativistic limit and hence seek spin-averaged masses. Details of the calculation are presented in the Appendices; the results for  spin averaged $S$-wave and $P$-wave charmonia and the spin averaged hybrid masses are shown in Fig. \ref{fig:Hmasses}. We remark that the method for spin-averaging hybrid multiplets is not obvious; here we simply take these values as the masses of the spin zero member of the multiplet. This procedure is justified in Ref. \cite{Lebed:2017xih}. The figure shows two sets of lattice charmonia masses obtained with light quark masses that correspond to pion masses of 400 MeV (dark boxes) and 236 MeV (light boxes). Model results have been centered on the lattice $\eta_c$--$J/\psi$ multiplet in the first column. The resulting spin-averaged $H_2$ and $H_3$ hybrid mass predictions are high by 100 -- 200 MeV. Somewhat surprisingly, given that no parameters were adjusted, the other mass predictions agree rather well with the more recent lattice field spectrum.  We find these initial results encouraging and look forward to performing more detailed computations. In the meantime, we adopt the hybrid wavefunction width parameters and move on to hybrid decays.

\section{Hybrid Decay Model}
\label{sec:decay}

\begin{figure}[ht]
\includegraphics[width=0.3\linewidth]{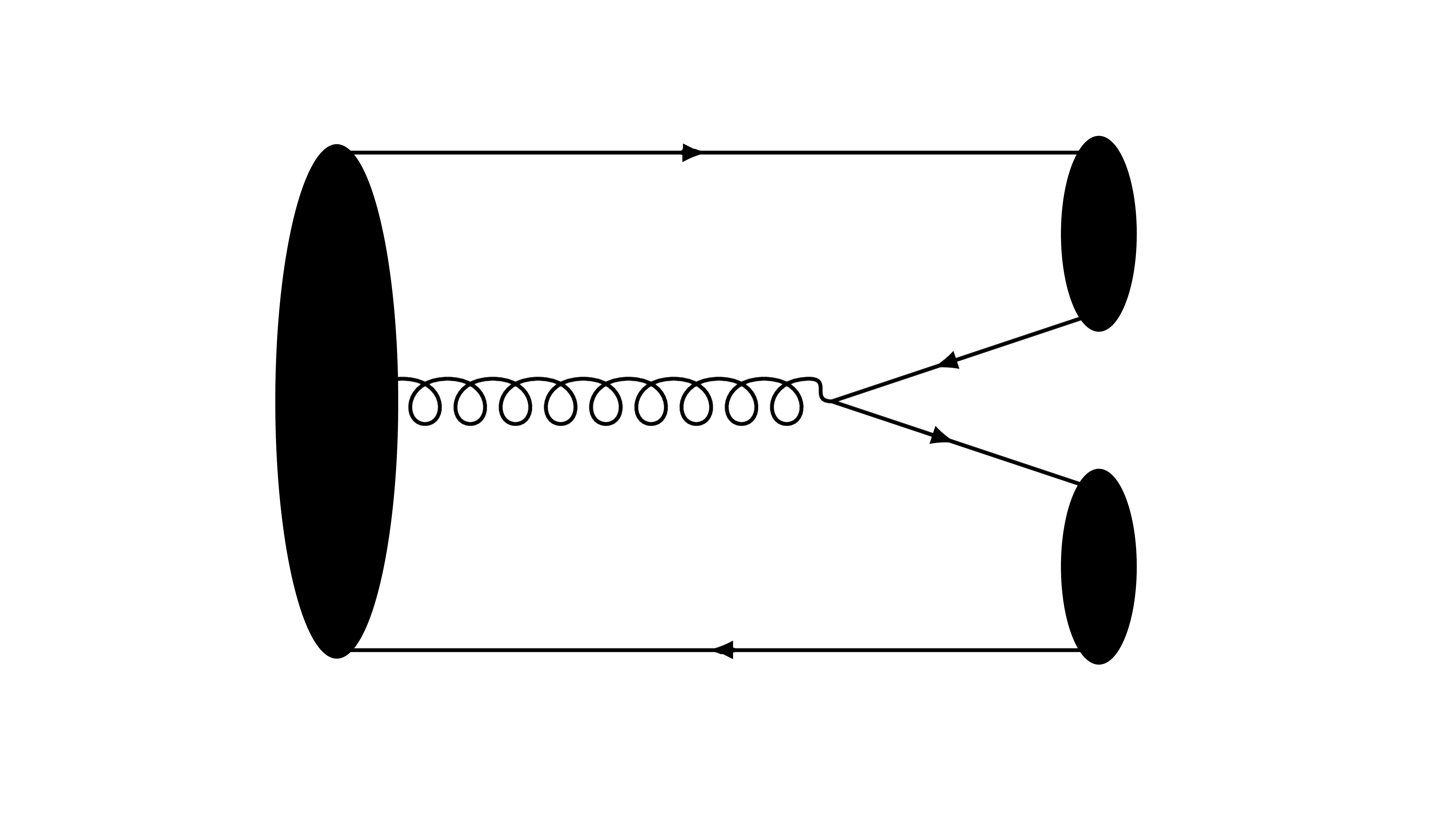}
\caption{Hybrid Decay Model.}
\label{fig:Hdecay}
\end{figure}

It is feasible and perhaps obligatory to consider strong hybrid decays once a reasonably robust model of hybrid structure is in hand. The first issue is determining the decay operator. Possibilities include absorbing the valence gluon into a quark or the Coulomb operator and then pair producing via a valence-valence-transverse tri-gluon vertex. This is rather elaborate and is suppressed by two powers of a mass scale (that is roughly 1 GeV). We choose, instead, to adopt the simplest and leading operator, which is quark pair production via gluon dissociation (Fig. \ref{fig:Hdecay}).

The calculation will be made using the hybrid wavefunctions found in the previous section. In a similar fashion, we also employ SHO wavefunctions for the final state mesons. Masses for the hybrids are taken from the lattice \cite{Liu:2012ze,Cheung:2016bym} and measured masses are used for all mesons. Details are provided in the Appendix along with a description of the $P$-wave mixing angles used.

The decay model contains two selections rules: (i) spin zero hybrids do not decay to spin zero mesons, (ii) TE hybrids do not decay to mesons with identical spatial wavefunctions. The former follows from the ${}3S_1$ quantum numbers of the vertex, while the latter is proved in the Appendix. These selection rules are typical of hybrid decay models, and there is reason to believe that they are generally true\cite{Page:1996rj}.

The partial widths obtained in the model generally follow the pattern of ``${}^3S_1$" models, namely the leading partial wave is much larger than subleading ones. It is interesting that exceptions do exist; for example $H_2(1+-) \to  D^*\bar D$ has an S/D ratio of 0.16. It is possible that this unusual circumstance can aid in the eventual identification of hybrid candidates. More generally, partial wave ratios tend to be closer to unity for decay models in which the vertex carries ${}^3P_0$ quantum numbers. In fact this can be used to distinguish strong decay models in the case of canonical meson decays\cite{Ackleh:1996yt}.

% 1+-(H2) DECAYS  mA =    4.3439998626708984     
% D* D   q=  0.97719825513494407     
%  threshold:    3.8789999485015869     
%  1-- 0-+ S=1,L=0    1
%  1-- 0-+ S=1,L=2    6
%
%but reversed for Ds*Ds
% Ds* Ds   q=  0.81651065649932486     
%  1-- 0-+ S=1,L=0    47
%  1-- 0-+ S=1,L=2    6

The major open issue in the current work is the selection of the remaining parameter, the quark-gluon coupling  $g$. In contrast to all other model parameters, which have their values fixed by external data or by direct computation, the quark-gluon coupling is unconstrained. While it may be natural to assign

\be
g = \sqrt{4\pi a_S}
\ee
this is not necessarily accurate. The parameter $a_S$ results from a fit to either the meson spectrum or lattice Wilson loop computations of the adiabatic interquark potential. It therefore parameterises an effective description of  nonperturbative gluodynamics. It is possible, for example, that $a_S$ is approximated by the string splitting\cite{LW} 

\be
a_S \approx  \frac{\pi}{12}
\ee
at intermediate scales (say, 1/2 fm) and therefore has little direct contact with $g$.
%and that the perturbative $\alpha_S = g^2(\mu_R)/4\pi$ is only valid at very short distances,
% $\mu_R \sim 10$ 1/fm. 
A third alternative is that one should use the QCD coupling at the scale of hybrid mesons, hence 

\be
g \approx g_{QCD}(4.5\ \textrm{GeV}) \approx 1.6.
\ee 
%alpha(2.5) ~ 0.28
Alternatively, setting $g = g_{QCD}$ at the average gluon momentum gives $g \approx 2.5$.

The natural way to resolve the problem is to set the coupling from a well-established hybrid decay width. Unfortunately, no lattice field computations of open flavour heavy hybrid decay have been made. 

Turning to experiment, the quantum number exotic state $R_{c0}(4240)$ is reported in the Review of Particle Properties\cite{rpp}. The claimed quantum numbers are $J^{PC} = 0^{--}$, which are forbidden to fermion-antifermion systems. The state is seen to decay to $\pi^-\psi(2S)$ with a total width of $\Gamma = 220$ MeV with large errors. Note that the decay mode implies that the $R_{c0}$ is also flavour-exotic. A $0^{--}$ hybrid can be accommodated within the current model, but is a TM hybrid, and therefore  heavier than the 4.0 -- 4.5 GeV masses seen in the lightest multiplets. The poor agreement with model expectations, the exotic flavour content, and the closed flavour decay mode make it impossible to use this state for setting the coupling.

The $\psi(4260)$ (previously known as the $Y(4260)$) is a longstanding charmonium hybrid candidate. Its mass is reported as $4320(6)$ MeV and width as $55(19)$ MeV\cite{rpp}. The state is seen to decay to $\pi\pi J/\psi$, either directly, or via resonances as $f_0 J/\psi$ or $Z_c(3900)\pi$. Since the $Y$ does not fit with well-established quark model expectations\cite{Barnes:2005pb} it has been suggested that it is a hybrid candidate, with a natural identification with $H_1(1^{--})$. The lattice predictions for the mass of this state are 4285(14) MeV at $m_\pi=400$ MeV or 4411(17) MeV at $m_\pi=236$ MeV. The former value looks promising, so it is unfortunate that decreasing the pion mass towards its physical value worsens the agreement. This appears to be a general feature of the computation -- with all higher mass mesons shifting upwards in mass by 100 -- 200 MeV (these mesons have an excitation energy of approximately 1400 MeV; those with an excitation energy of approximately 900 MeV shift upwards by tens of MeV). An additional possible problem is that the closed flavour decay $H_1(1^{--}) \to \pi\pi J/\psi$ must be accompanied by a spin flip, which is generally suppressed. Countering this is the existence of anomalies in the hadronic transitions $\Upsilon(nS) \to \pi\pi \Upsilon(n'S)$ also imply a failure of naive expectations with respect to ``spin flip". This can happen if intermediate states such as $B^*\bar B$ can propagate for a long time and effectively decorrelate the heavy quark spin. Indeed, the observation of the $Y$ in $Z_c(3900)\pi$ indicates that such a mechanism could be relevant here. Thus, although it is tempting, it is not prudent to set the coupling by comparison with the $\psi(4260)$.

In view of the lack of external information on the value of $g$, we proceed with the initial impulse and set $g = \sqrt{4\pi a_S} \approx 2.73$ and bear in mind that the results may require substantial modification pending the arrival of new information. Of course this modification amounts to simply scaling the results, so at least it is not difficult to make.

Table \ref{tab:widths} presents partial decay widths for the complete set of low-lying hybrid mesons to a variety of channels. In this case we take the hybrid masses to be those obtained on the lattice with $m_\pi = 400$ MeV. Although these are somewhat lighter than those obtained with more physical light quark masses, the calculation illustrates the effect of assumptions about the hybrid masses and might be relevant to nature if it is confirmed that the $\psi(4260)$ does indeed contain substantial valence gluonic degrees of freedom.

The table reveals that members of the $H_1$ multiplet are  very narrow and should appear as sharp peaks in their final states. In contrast, all other hybrids have typical hadronic widths. Of course, this statement relies on the assumed value of the quark-gluon coupling constant.  
It is evident that S+P decay modes dominant total widths when these channels are open. This repeats a rule of thumb found in previous hybrid decay models, such as Ref. \cite{Isgur:1985vy}.

% computed with a_S = 0.594, betas as in table V + Close-Swanson meson betas, 
% NB: it is often assumed that theta_mixing=0 to simplify the D1L/D1H computations. Thus 
% D1L approx 1P1 and D1H approx 3P1. In fact sine^2(th) =19% so it's not a terrible approx
\begin{table}[ht]
\caption{Hybrid Decay Widths (MeV), computed with the $m_\pi= 400$ MeV Hybrid Spectrum. Suppressed channels are denoted as follows: $-$ = quantum number, \o = threshold, 0 = selection rule, x = negligible.}
\begin{tabular}{l|ccccccccccccc|c}
\hline\hline
state & $D^*D$ & $D_0D$ & $D_{1L}D$ & $D_{1H}D$ & $D_2D$ & $D_0D^*$ & $D_{1L}D^*$ & $D_{1H}D^*$ & $D_2D^*$ & $D_s D_s^*$ & $D_{s0} D_s$ & $D_{s0} D_s^*$ & $D_{s1L} D_s$ & width\\
\hline
$1^{--}$ $(H_1)$ & x & - & \o & \o & \o & \o & \o & \o & \o & \o & - & \o & \o & x \\
$0^{-+}$ $(H_1)$ & x & \o & \o & \o & \o & - & \o & \o & \o & \o & \o & - & \o & x\\
$1^{-+}$ $(H_1)$ & x & - & \o & \o & \o & \o & \o & \o & \o & 6 & - & \o & \o & x\\
$2^{-+}$ $(H_1)$ & x &  x &  x & x & 6 &\o & \o &\o &\o & \o &  \o &  \o & \o & x \\
\hline
$1^{++}$ $(H_2)$ & x  & - & 3 & 11 & x & 4 & \o & \o & \o & \o & - & \o & \o & 18\\
$0^{+-}$ $(H_2)$ & x &  x & 21 & 3 & x  & 4 & \o & \o   & \o  & x  & x& \o & \o & 28 \\
$1^{+-}$ $(H_2)$ & x &  12 & 1  & 1 & x & \o  & \o  & \o  &\o  & 1  &  6  &\o  &\o & 21  \\
$2^{+-}$ $(H_2)$ & x &  15 &  3 &  2 & 3 &  2  &  \o &  \o &  \o &  x &  12 &  \o  &   \o & 37 \\
\hline
$0^{++}$ $(H_3)$ & - & - & 3 & 13 & - & 10 & 2 & 2 & \o & - & - & 3 & 1 & 44\\
$1^{+-}$ $(H_3)$ & x &  4 &  6  &  2  & 4  & 6  &  3 &   2  &  x  & x &  4  &   2  &   1 & 34 \\
\hline
$2^{++}$ $(H_4)$ &  x  &  x  &   0 &  17  & x  & 12  & 9 & 2 &  1 &  x &  x &  6 & 0 & 47\\
$1^{+-}$ $(H_4)$ &  x & 24  & 8 & 4 & x  &  10  &   1  &  1  & x &  x & 28 & 6 &  3 & 85\\
$2^{+-}$ $(H_4)$ & x &  x  & 26 &  6 & 1 &  20 &  5 &  5 & 2 &  x &  x &  12 &  13 & 90 \\
$3^{+-}$ $(H_4)$& x &  x &   0  &   x   &   10  & x   & x & 26 & 7 & x &  x  &   x & 0 & 43\\
\hline\hline
\end{tabular}
\label{tab:widths}
\end{table}

\begin{table}[ht]
\caption{Hybrid Decay Widths (MeV), computed with the $m_\pi= 236$ MeV Hybrid Spectrum. Suppressed channels are denoted as follows: $-$ = quantum number, \o = threshold, 0 = selection rule, x = negligible.}
\begin{tabular}{l|ccccccccccccc|c}
\hline\hline
state & $D^*D$ & $D_0D$ & $D_{1L}D$ & $D_{1H}D$ & $D_2D$ & $D_0D^*$ & $D_{1L}D^*$ & $D_{1H}D^*$ & $D_2D^*$ & $D_s D_s^*$ & $D_{s0} D_s$ & $D_{s0} D_s^*$ & $D_{s1L} D_s$ & width\\
\hline
$1^{--}$ $(H_1)$ & x & - & 8 & 33 & x & 25 & \o & \o & \o & x & - & \o & \o & 66\\
$0^{-+}$ $(H_1)$ & x & 76 & \o & \o & \o & - & \o & \o & \o & 34  & \o & \o & \o & 76\\
$1^{-+}$ $(H_1)$ & x & - & 28 & \o & \o & \o & \o & \o & \o & x  & - & \o & \o & 28\\
$2^{-+}$ $(H_1)$ & x & x  &  x & 1 & 20 &  x & 16 & 24 &\o & x & x & x & x & 60\\
\hline
$1^{++}$ $(H_2)$ & x  & - & 5 & 18 & x & 11 & 5 & x & \o & x & - & 4 & 2 & 45\\
$0^{+-}$ $(H_2)$ & x &  x & 30 & 5 & x  & 17 & x & \o   & \o  & x  & x & x & 1 & 53  \\
$1^{+-}$ $(H_2)$ & x & 16  & 4  & 3 & 5 & 4  & x  & \o  & \o  & x  & 16   & x  & x & 48  \\
$2^{+-}$ $(H_2)$ & x & 16  & 4  & 3  & 6 & 7  & 14  & 5 & 1  & x & 20  & 4   & 2 & 82  \\
\hline
$0^{++}$ $(H_3)$ & - & - & 4 & 15 & - & 15 & 7 & 21 & 14 & - & - & 14 & 6 & 96 \\
$1^{+-}$ $(H_3)$ & x & 4  & 6   & 3   & 4  & 9  & 11  & 16  & 5   & x & 5   & 8  & 5 & 76   \\
\hline
$2^{++}$ $(H_4)$ & x &  x  &   0 & 17   & x & 15  & 23   & 9 & 14  & x  &  x & 18 & 0 & 96\\
$1^{+-}$ $(H_4)$ & x  & 17  & 7 & 4 & x  & 12   & 2  & 4  & 3 & x  & 31 & 17 & 8 & 105  \\
$2^{+-}$ $(H_4)$ & x &  x  &  23  & 6  & 2 & 22  & 10  & 14  & 9 & x   &  x & 28 & 29 & 139  \\
$3^{+-}$ $(H_4)$&  x   &  x    &   0  &  x    &  9 & x   & x & 40 & 13 & x & x   & x   & 0 & 62\\
\hline\hline
\end{tabular}
\label{tab:widthsHeavy}
\end{table}

Table \ref{tab:widthsHeavy} presents predicted widths when the hybrid masses are fixed according to the lattice results with $m_\pi=236$ MeV\cite{Cheung:2016bym}. In this case more S+P channels open for the $H_1$ multiplet and these states attain typical charmonium widths. 
As an illustration of an application of these predictions, we show total hybrid widths for $m_\pi= 236$ MeV hybrid  masses in Table \ref{tab:totals}. Widths have been normalized by identifying the $H_1(1^{--})$ as the $\psi(4660)$. (It is worth stressing that this identification is not entirely satisfactory as corresponding lattice mass is 4411 MeV.) One observes that the predicted widths happen to be close to those obtained with our assumed coupling (Table \ref{tab:widthsHeavy}).

\begin{table}[ht]
\caption{Total Hybrid Widths (MeV). Normalized by assuming $H_1(1^{--}) \leftrightarrow \psi(4660)$.}
\begin{tabular}{llll|llll|ll|llll}
\hline\hline
\multicolumn{4}{c|}{$H_1$} & \multicolumn{4}{c|}{$H_2$} & \multicolumn{2}{c|}{$H_3$} & \multicolumn{4}{c}{$H_4$} \\
$1^{--}$ & $0^{-+}$ & $1^{-+}$ & $2^{-+}$ & $1^{++}$ & $0^{+-}$ & $1^{+-}$ & $2^{+-}$ & $0^{++}$ & $1^{+-}$ & $2^{++}$ & $1^{+-}$ & $2^{+-}$ & $3^{+-}$ \\
\hline
70 & 83 & 30 & 66 & 50 & 57 & 53 & 90 & 103 & 81 & 103 & 113 & 154 & 68 \\
\hline\hline
\end{tabular}
\label{tab:totals}
\end{table}

It is interesting to enquire into the feasibility of using decay characteristics as a signature of gluonic content. Table \ref{tab:comparison} shows partial decay widths for the cryptoexotic $H_1(1^{--})$ at an assumed mass of 4111 MeV and the $4{}^3S_1$ charmonium vector at the same mass. The latter decay rates have been computed with the ``3P0" model\cite{Barnes:2005pb}. Although the predicted total widths are similar, it is evident that the decay patterns are quite different; especially in the S+S channels and in the reversed roles played by $D_{1L}D$/$D_{1H}D$ and $D_2D$/ $D_0D^*$ decays.

\begin{table}[ht]
\caption{Cryptoexotic $H_1(1^{--})$ and $\psi(4{}^3S_1)$ Decay Modes (assuming masses of 4411 MeV).}
\begin{tabular}{l|cccccccccc|c}
\hline\hline
state & $DD$ & $D^*D$ & $D^*D^*$  & $D_{1L}D$ & $D_{1H}D$ & $D_2D$ & $D_0D^*$ & $D_sD_s$ & $D_s D_s^*$ & $D_s^*D_s^*$ & width\\
\hline
$H_1(1^{--})$ & 0 & 0.078 & 0 &  8 & 33 & 0.0035 & 25 & 0 & 0.2 & 0 & 66 \\
$\psi(4{}^3S_1)$ & 0.4 & 2.3 & 16 & 31 & 1 & 23 & 0 & 1.3 & 2.6 & 0.7 & 78 \\
\hline\hline
\end{tabular}
\label{tab:comparison}
\end{table}

\section{Conclusions}
\label{sec:conc}

We have developed a simple model of hybrid structure that  leverages guidance from QCD in Coulomb gauge and from lattice field theory computations. In particular we assume that an axial constituent gluon provides a reasonable facsimile of low lying gluonic degrees of freedom. Certainly, this is in agreement with recent lattice computations of the charmonium spectrum. Furthermore, when the interactions of the constituent gluon are sufficiently tuned (in particular, by including three-body interactions), it is possible to maintain consistency with the gluelump spectrum and the gluonic adiabatic energy surfaces. Using these observations as a starting point has led to a reasonable spin-averaged spectrum shown in Fig. \ref{fig:Hmasses} and lends confidence that a more detailed calculation should be able to provide an accurate representation of the lowest hybrid multiplets. It will be interesting to improve the spectrum computation and to include spin-splittings. Having a detailed structure model will also permit the computation of other observables of interest, such as radiative couplings.

Adopting a constituent gluon model for hybrid structure leads naturally to the assumption that quark pair production from the valence gluon describes strong hybrid meson decay, resurrecting an old model of Tanimoto. The model incorporates several familiar selection rules: spin zero hybrids do not decay to spin zero mesons, TE hybrids do not decay to identical mesons or mesons with identical spatial wavefunctions, and S+P wave decay modes are preferred. We note that the structure of the decay vertex satisfies the general conditions of Ref. \cite{Barnes:2007xu}, and therefore hybrid mesons will experience no relative mass shifts if spin-averaged hadron masses are assumed. 

With respect to the present work, the major remaining issue is the determination of the quark-gluon coupling. The most likely way forward is a direct lattice computation of an open flavour decay. Alternatively, the dependence of hybrid masses on the light sea quark mass evident in Table \ref{tab:spectrum} (the quoted lattice computations are unquenched) is an indication that relatively large couplings of these states to the continuum occur. The fact that the mass shifts are comparable to those of canonical charmonia is also an indication that the widths predicted here are of the correct magnitude.

\acknowledgments

Swanson acknowledges support by the U.S. Department of Energy under contract DE-SC0019232. Garcia-Tecocoatzi acknowledges financial support from Consejo Nacional de Ciencia y Tecnología, Mexico.

\appendix

\section{Parameters}

Quark model parameters used in the spectrum computation are as follows:

\be
a_S = 0.594 \qquad b = 0.16\ \textrm{GeV}^2 \qquad m_c = 1.6\ \textrm{GeV}\qquad m_g = 0.6\ \textrm{GeV} \qquad b_g = 6\  \textrm{GeV}.
\ee
The first three values are typical of potential quark models\cite{Swanson:1992ec}, while the latter two are obtained in a mean field computation of the gluonic vacuum structure reported in Ref. \cite{Szczepaniak:2001rg}. Additional quark masses required for the decay computation were taken to be $m_u = 0.33$ GeV and $m_s = 0.55$ GeV.

Hybrid masses are obtained from two lattice field computations of the charmonium spectrum, with hybrids identified by the strength of state coupling to operators with explicit gluonic fields.  These masses are displayed in Table \ref{tab:spectrum}.

\begin{table}[h]
\caption{Lattice hybrid charmonium spectra employed in the computation.}
\begin{tabular}{c|ccc}
\hline\hline
multiplet & $J^{PC}$ & mass (MeV)\cite{Liu:2012ze} & mass (MeV)\cite{Cheung:2016bym} \\
\hline
$H_1$ & $1^{--}$ & 4285(14) & 4411(17) \\
& $0^{-+}$ & 4195(13)       & 4279(18) \\
& $1^{-+}$ & 4217(16)       & 4310(23) \\
& $2^{-+}$ & 4334(17)       & 4456(21) \\ \hline
$H_2$ & $1^{++}$ & 4399(14) & 4470(25) \\
& $0^{+-}$ & 4386(09)       & 4437(27) \\
& $1^{+-}$ & 4344(38)       & 4438(23) \\
& $2^{+-}$ & 4395(40)       & 4502(18) \\ \hline
$H_3$ & $0^{++}$ & 4472(30) & 4591(46) \\
& $1^{+-}$ & 4497(19)       & 4571(27) \\ \hline
$H_4$ & $2^{++}$ & 4492(21) & 4623(32) \\
& $1^{+-}$ & 4497(39)       & 4665(53) \\
& $2^{+-}$ & 4509(18)       & 4631(26) \\
& $3^{+-}$ & 4548(22)       & 4644(34) \\
\hline\hline
\end{tabular}
\label{tab:spectrum}
\end{table}

The wavefunction width parameters used in the decay calculation are given in Table \ref{tab:betas}. These are drawn from the cited references for mesons and are obtained from our variational calculation for hybrids.

\begin{table}[h]
\caption{Gaussian width parameters (MeV).}
\begin{tabular}{l|cc|cc|c}
\hline\hline
state & $c\bar u$\cite{Close:2005se} & $c\bar c$\cite{Close:2005se} & $c\bar u$\cite{Godfrey:2015dva} & $c\bar s$\cite{Godfrey:2015dva} & hybrid $(\alpha,\beta)$ \\
\hline
     Ps  &   430  &   710 &  600 &  650 &  \\
     V   &   370  &   660 &  520 &  560 &  \\
     $^1P_1$ & 330&   500 &  470 & 490 &   \\
     $^3P_J$ & 320 & 490 &  430& 460 &   \\
     $^3D_J$ & 300 & 450 & 410 & 400 &  \\
     $^1D_2$ & 300 & 450 & 430 & 440 &  \\
     Ps(2S)  & 310  & 480 & 450 &  480  &  \\
     V(2S)   & 300  & 480  & 430 &  460  &   \\
     Ps(3S)  & 280  &  410 & 410 &  420  &  \\
     V(3S)   & 270 &  410  & 400 &  420   &  \\
\hline
  $H_1$  &          &        &      &        &  415, 680 \\
  $H_2$  &          &        &      &        &  405, 630 \\
  $H_3$  &          &        &      &        &  405, 600 \\
  $H_4$  &          &        &      &        &  405, 620 \\
\hline\hline
\end{tabular}
\label{tab:betas}
\end{table}

Mixed open flavour $P$-wave states are defined via a rotation from quark model states with fixed charge conjugation as follows:

\begin{eqnarray}
D_{1L} &=& \phantom{-}\cos(\theta) \, |^1P_1\rangle + \sin(\theta)\, |^3P_1\rangle \nonumber \\
D_{1H} &=& -\sin(\theta) \, |^1P_1\rangle + \cos(\theta)\, |^3P_1\rangle,
\label{eq:mixing}
\end{eqnarray}
with similar relationships for the $D_{s1}$ states. The mixing angle arises from a tensor spin-dependent interaction in the potential quark model. A specific model computation\cite{Godfrey:2015dva} yields the results

\be 
\theta_{1P}(cu) = -25.7^\circ \quad 
\theta_{2P}(cu) = -29.4^\circ \quad
\theta_{1P}(cs) = -37.5^\circ \quad 
\theta_{2P}(cs) = -30.4^\circ 
\label{eq:theta}
\ee
By way of comparison, these angle are $\theta_{1P} = -54.7$ or $35.3$  degrees in the heavy quark limit.
%\cite{Lakhina:2006fy}

\section{Hybrid Wavefunctions}

We seek to estimate spin averaged hybrid masses with the variational Ansatz of Eqs. \ref{eq:PSI} and \ref{eq:ansatz}. In the case of equal quark masses the kinetic energy operators are purely radial and the angular integrals of Eq. \ref{eq:var} can be performed. The result is diagonal in the quantum numbers and of the form (delta functions in quantum numbers are suppressed in the following)

\be
\langle K \rangle = \int \frac{k^2 dk}{(2\pi)^3} \, \frac{q^2 dq}{(2\pi)^3}\, |\varphi_{j_g}(k)|^2 \, |\varphi_\ell(q)|^2\, \left[ 2m_c + \frac{q^2}{m_c} + \frac{k^2}{4 m_c} + \frac{1}{2} \left(\omega(k) + \frac{k^2}{\omega(k)}\right)\right].
\ee

The quark-antiquark interaction gives the contribution
\begin{eqnarray}
\langle V_{q\bar q} \rangle &=& +\frac{1}{6}\int \frac{d^3q}{(2\pi)^3}\, \frac{d^3Q}{(2\pi)^3} \, \varphi_\ell(q) Y_{\ell m_\ell}(\hat q)\, \varphi^*_\ell(q-Q) Y^*_{\ell m_\ell}(\widehat{q-Q}) \, V(Q) \\ \nonumber
 &=& +\frac{1}{6} \int x^2 dx\, |\varphi_\ell(x)|^2 V(x).
\end{eqnarray}
Here $V$ is defined via $K^{AB}(r) = \delta^{AB} V(r)$ (see Eq. \ref{eq:K}).

It is worth remarking that, although the Fourier transform of a linear function does not exist, it is easy to convince oneself that this should be irrelevant when integrating against strongly convergent functions such as Gaussians. Indeed, regulating the linear potential yields a Fourier transform

\be
{\cal F}(x) = -\frac{8\pi}{(k^2+\epsilon^2)^3} (k^2-3\epsilon^2),
\ee
which gives correct results in integrals like above when the limit as $\epsilon$ goes to zero is taken.

The quark-gluon and antiquark-gluon contributions to the energy contain factors of $\frac{1}{2}[\sqrt{\omega(k+Q)/\omega(k)}+\sqrt{\omega(k)/\omega(k+Q)}]$. This is equal to unity for large $k$ and for small $k$ and $k+Q$. This leaves a small and inconsequential region where this factor deviates from unity, hence we simply set this equal to one (doing so permits doing the relevant integrals analytically). Explicit expressions are

\begin{eqnarray}
\langle H_{qg}\rangle = \langle H_{\bar q g}\rangle &=& -\frac{3}{2}\int \frac{d^3k}{(2\pi)^3} \frac{d^3q}{(2\pi)^3} \frac{d^3Q}{(2\pi)^3} \, V(Q)\, \varphi_{j_g}(k)  \varphi_\ell(q) Y_{\ell m_\ell}(\hat q) \varphi^*_{j_g'}(k+Q) \varphi_{\ell'}^*(q-Q/2) Y^*_{\ell' m_\ell'}(\widehat{q-Q/2})\, \cdot \\ \nonumber
 &&  \frac{3}{4\pi}\, D^{1*}_{b0}(\hat k) D^1_{b'0}(\widehat{k+Q}) \, \langle 1 m_g 1 i | 1 b \rangle \, \langle 1 m_g' 1 i| 1 b'\rangle \cdot \\ \nonumber
&& \langle \ell m_\ell 1 m_g| LM_L\rangle \, \langle \ell' m'_\ell 1 m_g'| L' M_L'\rangle \, \langle S M_S L M_L |JM\rangle \, \langle S M_S L' M_L'|J' M'\rangle.
\end{eqnarray}
The integrals over $k$ and $q$ can be performed leaving relatively simple functions of $Q$. Finally, the integral over $Q$ can also be evaluated analytically. Intermediate results for the relevant cases of $\ell=0$ and $\ell=1$ are 

\be
\langle V_{qg}\rangle_{\ell=0} = -\frac{3}{2}\int \frac{Q^2 dQ}{2\pi^2} \, V(Q) {\rm e}^{-Q^2/16 \alpha^2} \, {\rm e}^{-Q^2/4 \beta^2} \left( 1 - \frac{Q^2}{6 \beta^2}\right).
\ee
and
\be
\langle V_{qg}\rangle_{\ell=1} = -\frac{3}{2}\int \frac{Q^2 dQ}{2\pi^2} \, V(Q) {\rm e}^{-Q^2/16 \alpha^2} \, {\rm e}^{-Q^2/4 \beta^2} \left( 1 - \frac{Q^2}{24 \alpha^2} - \frac{Q^2}{6\beta^2}  +\frac{Q^4}{480 \alpha^2\beta^2} f_L\right),
\ee
with $f_0 = 0$, $f_1 = 5$, and $f_2 = 3$. In all cases $J=J'$ emerges as expected.

\section{Hybrid Decay}

We consider the decay $H\to B(K)C(-K)$ where $K$ is the momentum of the meson $B$. With this notation, the expression for the decay amplitude reads

\begin{eqnarray}
{\cal A}(K) &=& -\frac{2}{3}g \int \frac{d^3k}{(2\pi)^3}\frac{d^3q}{(2\pi)^3}\,
\Psi_{j_g,\ell}({\bf k}, {\bf q})\, \sqrt{\frac{2 j_g +1}{4\pi}} \, D_{m_g\mu}^{j_g*}(\hat k) \, \chi^{(\xi)}_{\mu,\lambda} \cdot\nn \\
&& \langle \frac{1}{2} m \frac{1}{2} \bar{m} | S M_{S} \rangle \,
\langle \ell m_\ell, j_g m_g| L M_L\rangle \, 
\langle S M_S, L M_{L} | J M \rangle \, \cdot \nn \\
&& \frac{1}{\sqrt{2\omega(k)}} \chi_s^\dagger {\bm \sigma} \tilde{\chi}_{s'} \cdot {\bm \epsilon}({\bm k},\lambda) \cdot \nn \\
&& \phi_B^*(\q-\k/2 -\frac{m_c}{m_q+m_c}{\bm K}) \, \langle \frac{1}{2}m \frac{1}{2} s'| S_B M_{SB}\rangle\, \langle S_B M_{SB} L_B M_{LB}|J_B M_B\rangle \cdot \nn \\
&& \phi_C^*(\q+\k/2 -\frac{m_c}{m_q+m_c}{\bm K}) \, \langle \frac{1}{2}s \frac{1}{2} \bar{m}| S_C M_{SC}\rangle\, \langle S_C M_{SC} L_C M_{LC}|J_CM_C\rangle.
\end{eqnarray}
This can be simplified by noting that 
\begin{align}
 \chi_{s}^{\dagger} { \boldsymbol  \sigma } \tilde{\chi}_{s'}   {\boldsymbol \epsilon}_{{\bf k},\lambda}=
- \sqrt{2}^{|s+s'|}D^{1}_{s+s^{'},\lambda}({\bf k}),
\label{wigner}
\end{align}
combining the two Wigner functions, and performing the sums over $\lambda$ and $\lambda'$. For a TE gluon with $j_g=1$ one obtains 

\be
\sqrt{2}^{|s+s'|} \, (-)^{s+s'} \, \langle 1 m_g 1 -(s+s')| 1b\rangle D^{1*}_{b0}(\hat k).
\ee
The integral over $\hat k$ depends on the product $\phi_B\phi_C$, which is even under $k\to -k$ if the wavefunctions are identical. Thus we derive the selection rule that TE hybrids cannot decay to pairs of identical mesons.

Decay rates were computed by performing sums in amplitudes analytically with Mathematica. The resulting six-dimensional integrals were then evaluated numerically and Clebsched to form partial widths of good total spin and angular momentum. It was crucial to set the final meson momentum along the $\hat z$ axis to make the latter steps feasible.

\begin{figure}[ht]
\includegraphics[width=0.75\linewidth]{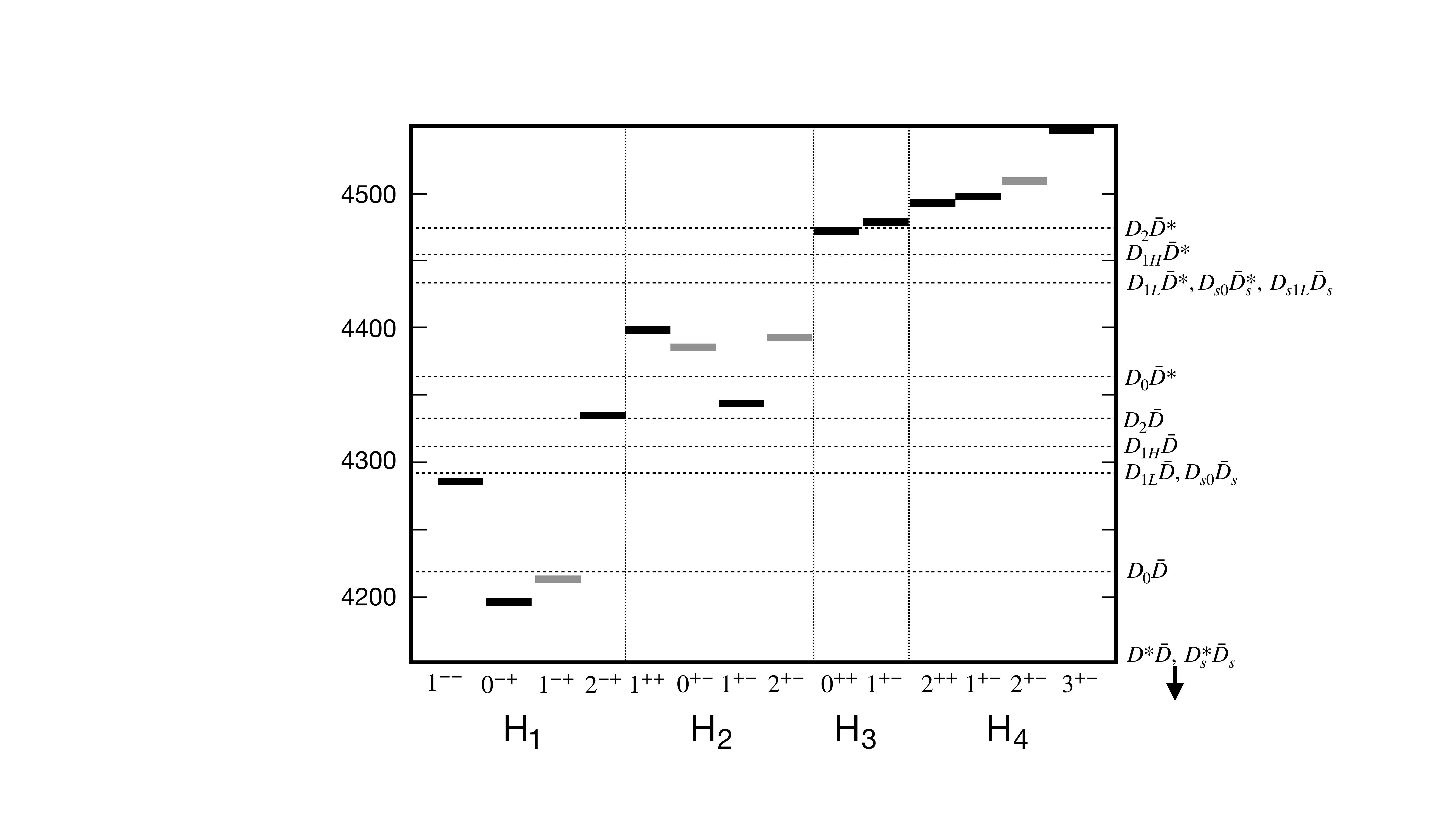}
\caption{Lattice Hybrid Masses ($m_\pi = 400$ MeV)  and Thresholds. Quantum number-exotic hybrids are denoted with grey boxes.}
\label{fig:Hdetail}
\end{figure}

Finally, for large $g$, it is the availability of phase space that determines the width of most hybrids. Relevant thresholds for the $m_\pi=400$ MeV case are shown in Fig. \ref{fig:Hdetail}.

\begin{thebibliography}{}

\bibitem{Dobbs}
S. Dobbs [GlueX],
% Searches for Exotic Hadrons at GlueX
[arXiv:1908.09711 [nucl-ex]].

%\cite{Lutz:2009ff}
\bibitem{Lutz:2009ff}
M.~Lutz \textit{et al.} [PANDA],
%``Physics Performance Report for PANDA: Strong Interaction Studies with Antiprotons,''
[arXiv:0903.3905 [hep-ex]].
%586 citations counted in INSPIRE as of 15 May 2020

\bibitem{bags}
T. Barnes, Caltech Ph.D. thesis (1977);
T. Barnes, Nucl. Phys. B158, 171 (1979);
P. Hasenfratz, R.R. Horgan, J. Kuti and J.M. Richard, Phys. Lett. {\bf 95B}, 299 (1980);
T. Barnes and F.E. Close, Phys. Lett. 116B, 365 (1982);
M. Chanowitz and S.R. Sharpe, Nucl. Phys. B222, 211 (1983);
T. Barnes, F.E. Close, and F. de Viron, Nucl. Phys. B224, 241 (1983);
M. Flensburg, C. Peterson, and L. Sk\"old, Z. Phys. C22, 293 (1984).


\bibitem{HM}
D. Horn and J. Mandula, Phys. Rev. D {\bf 17}, 898 (1978);
A. LeYaouanc, L. Oliver, O. P\`ene, J.-C. Raynal and S. Ono, Z. Phys. \textbf{C28}, 309
(1985). % con glue model





\bibitem{string}
R.~Giles and S.-H~H.~Tye,
%``The Psi Spectroscopy Of A Charm String,''
Phys.\ Rev.\ Lett.\  {\bf 37}, 1175 (1976);
W. Buchm\"{u}ller and S.-H. H. Tye, Phys. Rev. Lett. {\bf 44}, 850 (1980);
  T.~J.~Allen, M.~G.~Olsson and S.~Veseli,
  %``Excited glue and the vibrating flux tube,''
  Phys.\ Lett.\ B {\bf 434}, 110 (1998);
  %[hep-ph/9804452].
  %%CITATION = HEP-PH/9804452;%%
M.~L\"{u}scher and P.~Weisz,
%``Quark confinement and the bosonic string,''
JHEP {\bf 0207}, 049 (2002).
%[arXiv:hep-lat/0207003].
%%CITATION = HEP-LAT 0207003;%%


\bibitem{flux}
N. Isgur and J. Paton, Phys. Lett. {\bf 124B}, 247 (1983);
N. Isgur and J. Paton, Phys. Rev. D {\bf 31}, 2910 (1985);
J. Merlin and J. Paton, J. Phys. {\bf G11}, 439 (1985);
J. Merlin and J. Paton, Phys. Rev. {\bf D35}, 1668 (1987);
 S.~Capstick and P.~R.~Page,
  %``Constructing hybrid baryons with flux tubes,''
  Phys.\ Rev.\ D {\bf 60}, 111501 (1999);
  %[nucl-th/9904041].
  %%CITATION = NUCL-TH/9904041;%%
G.~Chiladze, A.~F.~Falk and A.~A.~Petrov,
%``Hybrid charmonium production in B decays,''
Phys. Rev. D \textbf{58}, 034013 (1998),
doi:10.1103/PhysRevD.58.034013
[arXiv:hep-ph/9804248 [hep-ph]].
%42 citations counted in INSPIRE as of 02 Jun 2020

%\cite{Meyer:2015eta}
\bibitem{Meyer:2015eta}
C.~Meyer and E.~Swanson,
%``Hybrid Mesons,''
Prog. Part. Nucl. Phys. \textbf{82}, 21-58 (2015)
doi:10.1016/j.ppnp.2015.03.001
[arXiv:1502.07276 [hep-ph]].
%99 citations counted in INSPIRE as of 14 May 2020

%\cite{Juge:1997nc}
\bibitem{Juge:1997nc}
K.~Juge, J.~Kuti and C.~Morningstar,
%``Gluon excitations of the static quark potential and the hybrid quarkonium spectrum,''
Nucl. Phys. B Proc. Suppl. \textbf{63}, 326-331 (1998)
doi:10.1016/S0920-5632(97)00759-7
[arXiv:hep-lat/9709131 [hep-lat]].
%134 citations counted in INSPIRE as of 15 May 2020


\bibitem{Juge:1999ie}
K.~Juge, J.~Kuti and C.~Morningstar,
%``Ab initio study of hybrid anti-b g b mesons,''
Phys. Rev. Lett. \textbf{82}, 4400-4403 (1999)
doi:10.1103/PhysRevLett.82.4400
[arXiv:hep-ph/9902336 [hep-ph]].
%130 citations counted in INSPIRE as of 13 May 2020

\bibitem{gluelump}
  C.~Michael,
  %``Adjoint Sources in Lattice Gauge Theory,''
  Nucl.\ Phys.\ B {\bf 259}, 58 (1985);
  %%CITATION = NUPHA,B259,58;%%
 N. Campbell, I. Jorysz and C. Michael, Phys. Lett. {\bf 167B}, 91 (1986);
 M. Foster and C. Michael [UKQCD Collaboration], Phys. Rev. D 59, 094509 (1999);
  %[arXiv:hep-lat/9811010].  SU(3) up to 24^3 x 48  
G. S. Bali and A. Pineda, Phys. Rev. D 69, 094001 (2004).
% [arXiv:hep-ph/0310130].


%\cite{Liu:2012ze}
\bibitem{Liu:2012ze} 
  L.~Liu {\it et al.} [Hadron Spectrum Collaboration],
  %``Excited and exotic charmonium spectroscopy from lattice QCD,''
  JHEP {\bf 1207}, 126 (2012)
  doi:10.1007/JHEP07(2012)126
  [arXiv:1204.5425 [hep-ph]].
  %%CITATION = doi:10.1007/JHEP07(2012)126;%%
  %208 citations counted in INSPIRE as of 31 May 2019

%\cite{Cheung:2016bym}
\bibitem{Cheung:2016bym}
G.~K.~Cheung \textit{et al.} [Hadron Spectrum],
%``Excited and exotic charmonium, $D_s$ and $D$ meson spectra for two light quark masses from lattice QCD,''
JHEP \textbf{12}, 089 (2016)
doi:10.1007/JHEP12(2016)089
[arXiv:1610.01073 [hep-lat]].
%45 citations counted in INSPIRE as of 14 May 2020

%\cite{Knechtli:2019bqx}
\bibitem{Knechtli:2019bqx}
F.~Knechtli,
%``Charmonium and Exotics from Lattice QCD,''
EPJ Web Conf. \textbf{202}, 01006 (2019)
doi:10.1051/epjconf/201920201006
[arXiv:1902.07079 [hep-lat]].
%2 citations counted in INSPIRE as of 15 May 2020

%\cite{Guo:2008yz}
\bibitem{Guo:2008yz}
P.~Guo, A.~P.~Szczepaniak, G.~Galata, A.~Vassallo and E.~Santopinto,
%``Heavy quarkonium hybrids from Coulomb gauge QCD,''
Phys. Rev. D \textbf{78}, 056003 (2008)
doi:10.1103/PhysRevD.78.056003
[arXiv:0807.2721 [hep-ph]].
%95 citations counted in INSPIRE as of 15 May 2020

%\cite{Berwein:2015vca}
\bibitem{Berwein:2015vca}
  M.~Berwein, N.~Brambilla, J.~Tarr\'us Castell\'a and A.~Vairo,
  %``Quarkonium Hybrids with Nonrelativistic Effective Field Theories,''
  Phys.\ Rev.\ D {\bf 92}, no. 11, 114019 (2015)
  doi:10.1103/PhysRevD.92.114019
  [arXiv:1510.04299 [hep-ph]].
  %%CITATION = doi:10.1103/PhysRevD.92.114019;%%
  %43 citations counted in INSPIRE as of 31 May 2019

%\cite{Brambilla:2018pyn}
\bibitem{Brambilla:2018pyn}
N.~Brambilla, W.~K.~Lai, J.~Segovia, J.~Tarr\'us Castell\'a and A.~Vairo,
%``Spin structure of heavy-quark hybrids,''
Phys. Rev. D \textbf{99}, no.1, 014017 (2019)
doi:10.1103/PhysRevD.99.014017
[arXiv:1805.07713 [hep-ph]].
%18 citations counted in INSPIRE as of 15 May 2020


%\cite{Dudek:2011bn}
\bibitem{Dudek:2011bn}
J.~J.~Dudek,
%``The lightest hybrid meson supermultiplet in QCD,''
Phys. Rev. D \textbf{84}, 074023 (2011)
doi:10.1103/PhysRevD.84.074023
[arXiv:1106.5515 [hep-ph]].
%107 citations counted in INSPIRE as of 15 May 2020

%\cite{Swanson:1998kx}
\bibitem{Swanson:1998kx} 
  E.~S.~Swanson and A.~P.~Szczepaniak,
  %``Heavy hybrids with constituent gluons,''
  Phys.\ Rev.\ D {\bf 59}, 014035 (1999)
  doi:10.1103/PhysRevD.59.014035
  [hep-ph/9804219].
  %%CITATION = doi:10.1103/PhysRevD.59.014035;%%
  %43 citations counted in INSPIRE as of 31 May 2019

%\cite{Szczepaniak:2006nx}
\bibitem{Szczepaniak:2006nx}
A.~P.~Szczepaniak and P.~Krupinski,
%``Energy spectrum of the low-lying gluon excitations in the Coulomb gauge,''
Phys. Rev. D \textbf{73}, 116002 (2006)
doi:10.1103/PhysRevD.73.116002
[arXiv:hep-ph/0604098 [hep-ph]].
%24 citations counted in INSPIRE as of 15 May 2020

%\cite{Guo:2007sm}
\bibitem{Guo:2007sm}
P.~Guo, A.~P.~Szczepaniak, G.~Galata, A.~Vassallo and E.~Santopinto,
%``Gluelump spectrum from Coulomb gauge QCD,''
Phys. Rev. D \textbf{77}, 056005 (2008)
doi:10.1103/PhysRevD.77.056005
[arXiv:0707.3156 [hep-ph]].
%47 citations counted in INSPIRE as of 15 May 2020


\bibitem{Tanimoto}
M. Tanimoto, 
%``Decay Patterns of qqg Hybriod Mesons,''
 Phys. Lett. {\bf 116B}, 198, (1982).

%\cite{Isgur:1984bm}
\bibitem{Isgur:1984bm}
N.~Isgur and J.~E.~Paton,
%``A Flux Tube Model for Hadrons in QCD,''
Phys. Rev. D \textbf{31}, 2910 (1985)
doi:10.1103/PhysRevD.31.2910
%758 citations counted in INSPIRE as of 15 May 2020

%\cite{Isgur:1985vy}
\bibitem{Isgur:1985vy}
N.~Isgur, R.~Kokoski and J.~Paton,
%``Gluonic Excitations of Mesons: Why They Are Missing and Where to Find Them,''
AIP Conf. Proc. \textbf{132}, 242-248 (1985);
doi:10.1103/PhysRevLett.54.869
%330 citations counted in INSPIRE as of 15 May 2020
%\cite{Close:1994hc}
F.~E.~Close and P.~R.~Page,
%``The Production and decay of hybrid mesons by flux tube breaking,''
Nucl. Phys. B \textbf{443}, 233-254 (1995)
doi:10.1016/0550-3213(95)00085-7
[arXiv:hep-ph/9411301 [hep-ph]].
%259 citations counted in INSPIRE as of 15 May 2020

%\cite{Swanson:1997wy}
\bibitem{Swanson:1997wy}
E.~S.~Swanson and A.~P.~Szczepaniak,
%``Hybrid decays,''
Phys. Rev. D \textbf{56}, 5692-5695 (1997)
doi:10.1103/PhysRevD.56.5692
[arXiv:hep-ph/9704434 [hep-ph]].
%33 citations counted in INSPIRE as of 15 May 2020

%\cite{Page:1998gz}
\bibitem{Page:1998gz} 
  P.~R.~Page, E.~S.~Swanson and A.~P.~Szczepaniak,
  %``Hybrid meson decay phenomenology,''
  Phys.\ Rev.\ D {\bf 59}, 034016 (1999)
  doi:10.1103/PhysRevD.59.034016
  [hep-ph/9808346].
  %%CITATION = doi:10.1103/PhysRevD.59.034016;%%
  %133 citations counted in INSPIRE as of 31 May 2019


\bibitem{Schwinger}
J.~Schwinger, Phys.\ Rev.\ {\bf 127}, 324 (1962). 

%\cite{Christ:1980ku}
\bibitem{Christ:1980ku}
N.~H.~Christ and T.~D.~Lee,
%``Operator Ordering And Feynman Rules In Gauge Theories,''
Phys.\ Rev.\ D {\bf 22}, 939 (1980).
%%CITATION = PHRVA,D22,939;%%


%\bibitem{gribov}
%V.N. Gribov, Nucl. Phys. {\bf B139}, 1 (1978).

%\cite{Szczepaniak:2001rg}
\bibitem{Szczepaniak:2001rg}
A.~P.~Szczepaniak and E.~S.~Swanson,
%``Coulomb gauge QCD, confinement, and the constituent representation,''
Phys. Rev. D \textbf{65}, 025012 (2002)
doi:10.1103/PhysRevD.65.025012
[arXiv:hep-ph/0107078 [hep-ph]].
%283 citations counted in INSPIRE as of 13 May 2020

%\cite{Zwanziger:2002sh}
\bibitem{Zwanziger:2002sh}
D.~Zwanziger,
%``No confinement without Coulomb confinement,''
Phys. Rev. Lett. \textbf{90}, 102001 (2003)
doi:10.1103/PhysRevLett.90.102001
[arXiv:hep-lat/0209105 [hep-lat]].
%188 citations counted in INSPIRE as of 15 May 2020

%\cite{Greensite:2011zz}
\bibitem{Greensite:2011zz}
J.~Greensite,
{\sl An introduction to the confinement problem}, 
Lect. Notes Phys. \textbf{821}, 1-211 (2011)
doi:10.1007/978-3-642-14382-3.
%147 citations counted in INSPIRE as of 15 May 2020

%\cite{Swanson:1992ec}
\bibitem{Swanson:1992ec}
E.~S.~Swanson,
%``Intermeson potentials from the constituent quark model,''
Annals Phys. \textbf{220}, 73-133 (1992)
doi:10.1016/0003-4916(92)90327-I.
%105 citations counted in INSPIRE as of 15 May 2020

%\cite{Lebed:2017xih}
\bibitem{Lebed:2017xih}
R.~F.~Lebed and E.~S.~Swanson,
%``Heavy-Quark Hybrid Mass Splittings: Hyperfine and "Ultrafine",''
Few Body Syst. \textbf{59}, no.4, 53 (2018)
doi:10.1007/s00601-018-1376-9
[arXiv:1708.02679 [hep-ph]].
%6 citations counted in INSPIRE as of 14 May 2020

%\cite{Page:1996rj}
\bibitem{Page:1996rj}
P.~R.~Page,
%``Why hybrid meson coupling to two S wave mesons is suppressed,''
Phys. Lett. B \textbf{402}, 183-188 (1997)
doi:10.1016/S0370-2693(97)00438-3
[arXiv:hep-ph/9611375 [hep-ph]].
%56 citations counted in INSPIRE as of 16 May 2020

%\cite{Ackleh:1996yt}
\bibitem{Ackleh:1996yt}
E.~Ackleh, T.~Barnes and E.~Swanson,
%``On the mechanism of open flavor strong decays,''
Phys. Rev. D \textbf{54}, 6811-6829 (1996)
doi:10.1103/PhysRevD.54.6811
[arXiv:hep-ph/9604355 [hep-ph]].
%272 citations counted in INSPIRE as of 17 May 2020

\bibitem{LW}
%\cite{Luscher:2002qv}
M.~Luscher and P.~Weisz,
%``Quark confinement and the bosonic string,''
JHEP \textbf{07}, 049 (2002)
doi:10.1088/1126-6708/2002/07/049
[arXiv:hep-lat/0207003 [hep-lat]].
%282 citations counted in INSPIRE as of 16 May 2020

\bibitem{rpp}
M. Tanabashi \textit{et al.} [PDG], Phys. Rev. D \textbf{98}, 030001 (2018).

%\cite{Barnes:2005pb}
\bibitem{Barnes:2005pb}
T.~Barnes, S.~Godfrey and E.~Swanson,
%``Higher charmonia,''
Phys. Rev. D \textbf{72}, 054026 (2005)
doi:10.1103/PhysRevD.72.054026
[arXiv:hep-ph/0505002 [hep-ph]].
%595 citations counted in INSPIRE as of 16 May 2020


%\cite{Braaten:2014qka}
%\bibitem{Braaten:2014qka} 
%  E.~Braaten, C.~Langmack and D.~H.~Smith,
%  %``Born-Oppenheimer Approximation for the XYZ Mesons,''
%  Phys.\ Rev.\ D {\bf 90}, no. 1, 014044 (2014)
%  doi:10.1103/PhysRevD.90.014044
%  [arXiv:1402.0438 [hep-ph]].
%  %%CITATION = doi:10.1103/PhysRevD.90.014044;%%
%  %71 citations counted in INSPIRE as of 31 May 2019

%\cite{Close:2005se}
\bibitem{Close:2005se} 
  F.~E.~Close and E.~S.~Swanson,
  %``Dynamics and decay of heavy-light hadrons,''
  Phys.\ Rev.\ D {\bf 72}, 094004 (2005)
  doi:10.1103/PhysRevD.72.094004
  [hep-ph/0505206].
  %%CITATION = doi:10.1103/PhysRevD.72.094004;%%
  %174 citations counted in INSPIRE as of 31 May 2019

%\cite{Godfrey:2015dva}
\bibitem{Godfrey:2015dva}
S.~Godfrey and K.~Moats,
%``Properties of Excited Charm and Charm-Strange Mesons,''
Phys. Rev. D \textbf{93}, no.3, 034035 (2016)
doi:10.1103/PhysRevD.93.034035
[arXiv:1510.08305 [hep-ph]].
%80 citations counted in INSPIRE as of 15 May 2020

%\cite{Lakhina:2006fy}
%\bibitem{Lakhina:2006fy}
%O.~Lakhina and E.~S.~Swanson,
%``A Canonical Ds(2317)?,''
%Phys. Lett. B \textbf{650}, 159-165 (2007)
%doi:10.1016/j.physletb.2007.01.075
%[arXiv:hep-ph/0608011 [hep-ph]].
%69 citations counted in INSPIRE as of 16 May 2020
			
%\cite{Barnes:2007xu}
\bibitem{Barnes:2007xu}
T.~Barnes and E.~Swanson,
%``Hadron loops: General theorems and application to charmonium,''
Phys. Rev. C \textbf{77}, 055206 (2008)
doi:10.1103/PhysRevC.77.055206
[arXiv:0711.2080 [hep-ph]].
%104 citations counted in INSPIRE as of 17 May 2020


\end{thebibliography}
\end{document}